\documentclass[aps,prl,twocolumn,showpacs,floatfix]{revtex4} %,tightenlines,11pt,
\usepackage{graphicx}
\usepackage{dcolumn}
\usepackage{amsmath}
\usepackage{mathptmx}
\begin{document}

%\documentclass[graphicx,aps,multicol]{revtex4}
%\begin{document}

\title{Enhanced radiative strength in the quasi-continuum of $^{117}$Sn}

\author{U.~Agvaanluvsan$^{1,2}$, 
A.C. Larsen$^3$\footnote{Electronic address: a.c.larsen@fys.uio.no}, R.~Chankova$^{4,5}$, M.~Guttormsen$^3$, G.~E.~Mitchell$^{4,5}$, 
A.~Schiller$^6$, S.~Siem$^3$, and A.~Voinov$^6$}

\affiliation{$^1$Stanford University, Palo Alto, California 94305 USA}
\affiliation{$^2$MonAme Scientific Research Center, Ulaanbaatar, Mongolia}
\affiliation{$^3$Department of Physics, University of Oslo, N-0316 Oslo, Norway}
\affiliation{$^4$Department of Physics, North Carolina State University, Raleigh, NC 27695, USA}
\affiliation{$^5$Triangle Universities Nuclear Laboratory, Durham, NC 27708, USA}
\affiliation{$^6$Department of Physics, Ohio University, Athens, OH 45701, USA}

\date{\today}

\begin{abstract}
The radiative strength function of $^{117}$Sn has been measured up to the neutron separation energy using the ($^3$He,$^3$He$^\prime\gamma$) reaction. An increase in the slope of the strength function around $E_{\gamma}= 4.5$ MeV indicates the onset of a resonance-like structure, giving a significant enhancement of the radiative strength function compared to standard models in the energy region $4.5 \leq E_{\gamma} \leq 8.0$ MeV. For the first time, the functional form of this resonance-like structure has been measured in an odd tin nucleus below neutron threshold in the quasi-continuum region.  
\end{abstract}  

\maketitle

Average electromagnetic properties of atomic nuclei can be described by the radiative strength function (RSF). An improved knowledge of the RSF is important for many aspects of pure and applied nuclear physics, including calculations of nuclear reaction cross sections and nuclear reaction rates in extreme stellar environments. For transitions with electromagnetic character $X$, multipolarity $L$ and energy $E_\gamma$, the RSF is defined by~\cite{BE73}
\begin{equation}
f_{XL}(E_\gamma) = \langle\Gamma_{if}^{XL}(E_\gamma)\rangle\, \rho(E_i,J_i^\pi)/E_\gamma^{2L+1}.
\label{eq:partial}
\end{equation}
Here, $\langle\Gamma_{if}^{XL}\rangle$ is the mean 
value of the partial decay width between the initial and final states, and $\rho(E_i,J_i^\pi)$ is the level density for the initial excitation energy $E_i$ and spin/parity $J_i^\pi$. For the dominating dipole radiation, the RSF is given by $f(E_\gamma) = f_{E1}(E_\gamma) + f_{M1}(E_\gamma)$. 

The RSF reveals essential information on nuclear structure. In particular, electric transitions between excited states in the nucleus are mostly influenced by the proton charge distribution, while for magnetic transitions the neutrons contribute as well due to their magnetic moments. Also the shape and softness of the nuclear surface are important factors for the nuclear response to electromagnetic radiation. 

The most dominant feature of the RSF is the giant electric dipole resonance (GEDR), which is centered around $E_{\gamma}= 15$~MeV. In a macroscopic picture, the GEDR is due to the nuclear charge, the protons, oscillating against the neutron cloud.
% and thus behaving as a dipole antenna. 
Other resonances such as the magnetic dipole spin-flip resonance and the electric quadrupole resonance have also been discovered, but are in general significantly smaller in magnitude than the GEDR and have less influence on the RSF~\cite{RIPL}. However, there are collective modes such as the so-called M1 scissors mode~\cite{LoIudice,Bohl} and the E1 skin oscillation mode~\cite{Isacker} that are small compared to the GEDR, but still large enough to appear above the GEDR tail. Such resonances can be of great importance for the nucleosynthesis in supernovae~\cite{Goriely}. 

Recently, a resonance-like structure in the RSF was observed in the $^{129 - 133}$Sn and the $^{133,134}$Sb isotopes using relativistic Coulomb excitation measurements in inverse kinematics~\cite{Adrich,Klimkiewicz_2}. This E1-type pygmy resonance was located at $\gamma$-ray energies around $8-10$ MeV, and was interpreted as excess neutrons oscillating against the core nucleons. The summed $B(\mathrm{E1})\!\uparrow$ strength of the pygmy resonance was found to be 3.2 and 1.9~$e^2\mathrm{fm}^2$ for $^{130,132}$Sn, respectively, which correspond to $\approx 7$\% and $\approx 4$\% of the classical Thomas-Reiche-Kuhn (TRK) sum rule. As these measurements are restricted to excitation energies above the neutron separation energy $S_n$, it is an open question whether additional strength may be found at lower energies. 

In fact, E1 transitions clustered in the $6.0 - 8.5$ MeV region of $^{116,124}$Sn have been found and studied by means of nuclear resonance fluorescence (NRF, photon scattering) experiments~\cite{Govaert}. The strength estimated from the $\gamma$-line intensities in these experiments was 0.204(25) $e^2$fm$^2$ and 0.345(43)~$e^2$fm$^2$ for $^{116,124}$Sn, respectively, corresponding to $\approx 0.4$\% and $\approx 0.6$\% of the TRK sum rule. Preliminary results on $^{112,124}$Sn from the HI$\gamma$S facility at TUNL~\cite{Anton} are consistent with the results of~\cite{Govaert}. In addition, various random-phase approximation calculations predict E1 strength in this mass and energy region~\cite{Paar,Sarchi}. The calculations indicate that a stronger fragmentation of the dipole strength is expected in exotic nuclei with a large mass-to-charge ratio compared to stable tin isotopes, and this has so far been supported by~\cite{Govaert, Adrich, Klimkiewicz_2,Anton}. New $^{117}$Sn($\gamma,$n) cross-section data~\cite{Hiro} together with $^{116}$Sn(n,$\gamma$) cross-section measurements and calculations~\cite{Goriely3} give further support to the presence of a pygmy resonance above neutron threshold also in $^{117}$Sn.

For tin isotopes, one might also expect the appearance of enhanced M1 strength since the proton Fermi surface is located right in between the $g_{7/2}$ and $g_{9/2}$ orbitals, while for the neutrons the $h_{11/2}$ and $h_{9/2}$ orbitals come into play. Thus, the $g_{7/2}~\leftrightarrow~g_{9/2}$ (protons) and $h_{11/2}~\leftrightarrow~h_{9/2}$ (neutrons) magnetic spin-flip transitions may show up as concentrated strength in the RSF. 
In proton inelastic scattering experiments on $^{120,124}$Sn with $E_p = 200$ MeV at very forward angles \cite{Djalali}, an M1 resonance centered at an excitation energy $E \approx 8.5$ MeV is observed. Recent ($\gamma, n$) experiments on $^{91,92,94}$Zr \cite{Utsunomiya} have revealed an enhanced M1 resonance at 9 MeV in these nuclei, with about 75\% more strength than predicted by systematics.

In this Letter, we present complementary measurements to the above-mentioned experiments for the tin isotope $^{117}$Sn. The Oslo method permits the simultaneous determination of the level density and the radiative strength function~\cite{Sch00a}. For both of these quantities the experimental results cover an energy region where there is little information available and data are difficult to obtain. The Oslo method, which is sensitive to radiative strength for $E_{\gamma}\lesssim S_n$, reveals the total intensity and the functional form of the RSF for $\gamma$-ray transitions in the quasi-continuum. 

The experiment was carried out at the Oslo Cyclotron Laboratory using a 38-MeV $^3$He beam with an average beam current of $\approx$ 1.5 nA impinging on a $^{117}$Sn target with thickness of 1.9 mg/cm$^{2}$. Particle-$\gamma$ coincidence events were detected using the CACTUS multidetector array. The reaction channel$^{117}$Sn($^3$He,$^3$He$^\prime\gamma$)$^{117}$Sn was selected using eight particle telescopes placed at 45$^\circ$ with respect to the beam direction. Each telescope consists of a Si $\Delta E$ and a Si(Li) $E$ detector with thicknesses 140 $\mu$m and 3000 $\mu$m, respectively. An array of 28 collimated $5\times5$ inches NaI $\gamma$-ray detectors with a total efficiency of $\approx$ 15\% at $E_{\gamma}=1.33$ MeV was used. In addition, one Ge detector
was applied in order to estimate the spin distribution and determine the
selectivity of the reaction. The typical spin range is $J\sim 2-6 \hbar$.

The energy of the ejectile is transformed into excitation energy using reaction kinematics. The particle--$\gamma$-ray coincidence spectra were unfolded with the NaI response function using the Compton subtraction method~\cite{Sch00a}. A subtraction procedure is adopted to extract the first-generation matrix $P(E,E_{\gamma})$ containing the primary $\gamma$-rays emitted from a given excitation energy $E$~\cite{Sch00a}. The matrix is expressed as the product of level density ($\rho$) and RSF ($f$):
\begin{equation}
P(E,E_\gamma)\propto\rho(E-E_\gamma)\,f(E_\gamma)\,E_\gamma^3.
\label{eq:ba}
\end{equation}
This factorization is justified for nuclear reactions leading to a compound state prior to the subsequent $\gamma$ decay. The RSF is only dependent on the $\gamma$ energy according to the generalized form of the Brink-Axel hypothesis~\cite{ Br55+Ax62}, which states that the GEDR and any other collective excitation mode built on excited states have the same properties as those built on the ground state. The functions $\rho$ and $f$ are obtained iteratively by a globalized 
fitting procedure~\cite{Sch00a}. In this Letter, we shall only focus on the RSF. There are two normalization parameters to be determined, namely the scaling ($B$) and the slope correction ($\alpha$) of the RSF according to the expression $B\exp (\alpha E_{\gamma})f(E_{\gamma})$~\cite{Sch00a}. 
\begin{figure}[hbt]
\includegraphics[totalheight=8.5cm]{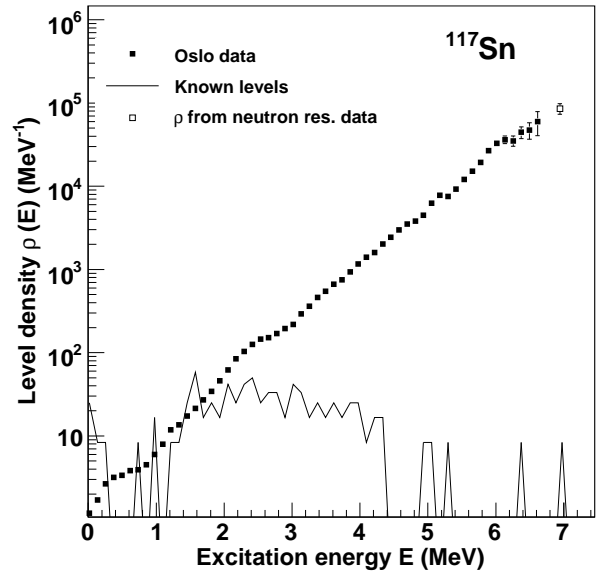}
\caption{Normalized level density of $^{117}$Sn.}
\label{fig:levdens}
\end{figure}

The $\alpha$ parameter is determined from the slope of the level density based on known low-lying discrete levels~\cite{ENSDF} and from the neutron resonance spacing $D$ at $S_n$ (see \cite{Sch00a} for details). We used the s- and p-wave resonance level spacings $D_0=(507\pm 60)$~eV and $D_1=(155\pm6)$~eV taken from~\cite{Mughabghab} to calculate the total level density at $S_n = 6.94$ MeV. Adopting the spin cut-off parameter $\sigma=4.44$~\cite{Egidy} and taking the average of the results obtained with $D_0$ and $D_1$, we obtained $\rho(S_n)= (8.55 \pm 1.24) \cdot 10^{4}$ MeV$^{-1}$. The normalized level density is shown in Fig.~\ref{fig:levdens}. 

\begin{figure}[hbt]
\includegraphics[totalheight=12.0cm]{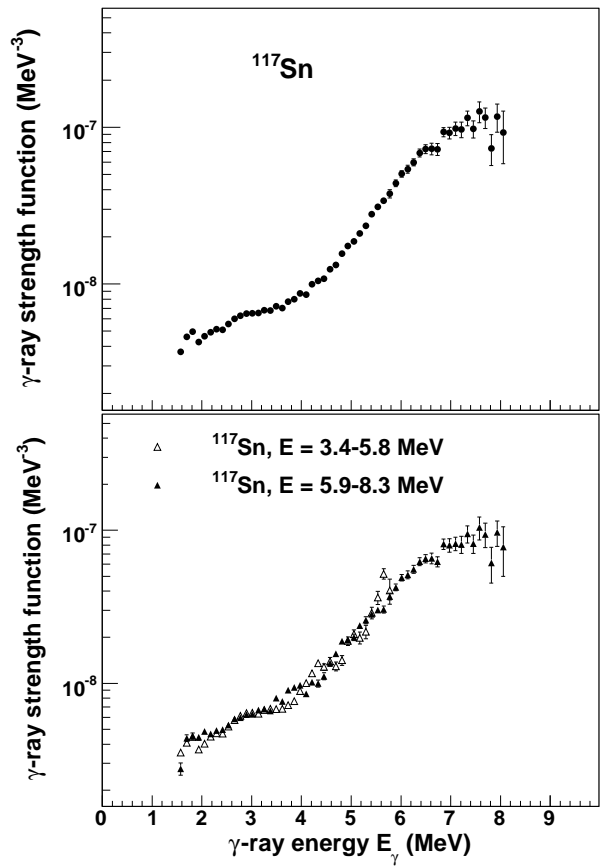}
\caption{Upper panel: Total radiative strength function of $^{117}$Sn. Lower panel: radiative strength function of $^{117}$Sn measured at two different excitation-energy regions.}
\label{lowhigh}
\end{figure}

The scaling parameter $B$ is determined using information on the average total radiative width $\langle\Gamma_\gamma\rangle$ at $S_n$ as described in Ref.~\cite{Voi01}. We normalize to $\langle\Gamma_\gamma\rangle$ = (53 $\pm 3)$~meV measured for s-wave neutron resonances~\cite{Mughabghab}. The normalized RSF of $^{117}$Sn is displayed in the upper panel of Fig.~\ref{lowhigh}. 

In order to test the assumption that the RSF of $^{117}$Sn is not dependent on excitation energy, we have deduced the RSF for two independent data sets of the experimental $P$ matrix. In the lower panel of Fig.~\ref{lowhigh} the resulting RSFs extracted for initial excitation energies in the intervals $3.4 \leq E \leq 5.8$ MeV and $5.9 \leq E \leq 8.3$ MeV are shown. The RSFs obtained from the two intervals are very similar, which indicates that the Brink-Axel hypothesis is indeed fulfilled in the excitation-energy region under consideration in this work.

We observe that the RSF exhibits an abrupt change in slope at $E_{\gamma}\approx 4.5$ MeV. Since there are strong indications of a pygmy resonance that could be due to skin oscillations in the even-even stable Sn isotopes, a similar phenomenon should be present in an odd-$A$ Sn nucleus with neutron excess as well. To investigate this further, we have compared our data to some of the available model predictions of the RSF. We have chosen two approaches: one where the strongest component, the E1 contribution from the GEDR, is assumed to follow a generalized Lorentzian (GLO)~\cite{RIPL,Kop93}, and one where the E1 strength is taken from microscopic calculations based on the quasi-particle random phase approximation (QRPA)~\cite{Goriely2}. The standard Lorentzian (SLO) model~\cite{RIPL}, which has been very successful in describing photonuclear cross-section data in the vicinity of the GEDR peak, has not been applied due to its tendency to overestimate average radiative widths and capture cross sections (see Ref.~\cite{RIPL} and references therein). This is seen to be the case also for $^{117}$Sn~\cite{Goriely3}, and we therefore deem that this model is not adequate to describe the experimental RSF below neutron threshold.  

\begin{figure}[htb]
\includegraphics[totalheight=12.0cm]{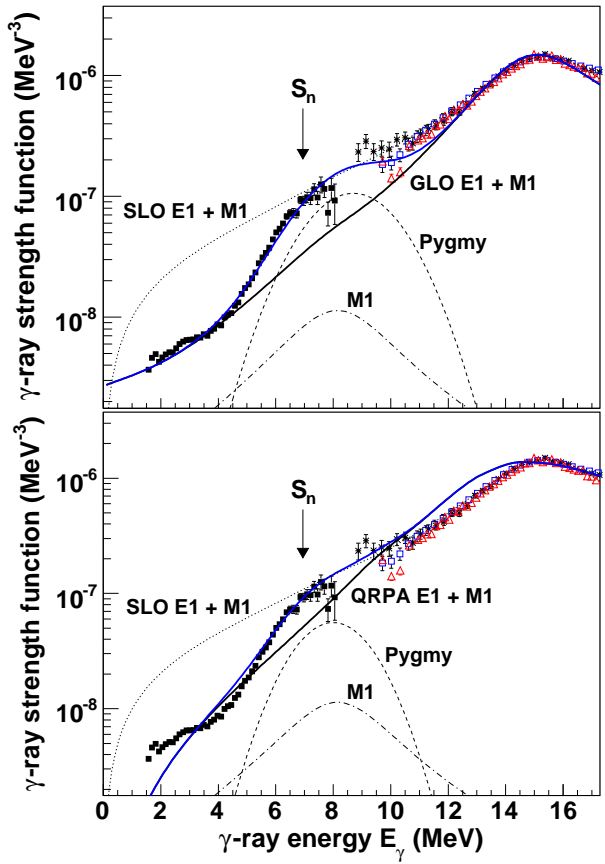} %Sn117_twomodels.eps
\caption{[Color online] Upper panel: Radiative strength function of $^{117}$Sn from the present work (black squares), and from ($\gamma,x$) photonuclear reactions (red open triangles~\cite{Fultz69}, stars~\cite{Lep74}, and blue open squares~\cite{Varlamov}). The sum of the GLO E1 strength and the M1 spin-flip resonance is shown as a solid line. The GLO E1 strength, the M1 spin-flip resonance (dashed-dotted line), and a Gaussian parameterization of the pygmy resonance (dashed line) are added to get a best fit (thick, blue solid line) to the data. The sum of the SLO E1 strength and the M1 strength is shown for comparison (dotted line). Lower panel: same as in the upper panel except that a QRPA E1 strength is used.}
\label{rsfboth}
\end{figure}

For the GLO approach we used experimental Lorentzian parameters taken from \cite{RIPL}. In the model calculations, we have treated the temperature of the final states $T_f$ as a free parameter. The best result was obtained using a constant temperature $T_f = 0.40$ MeV, for which we found a reasonable agreement with our data points in the region $1.5 \leq E_{\gamma} \leq 4.5$ MeV, and ($\gamma,x$) data~\cite{Fultz69,Lep74,Varlamov} around the GEDR peak ($13 \leq E_{\gamma} \leq 16$ MeV).  Using a constant temperature is consistent with the Brink-Axel hypothesis and our results (see Fig.~\ref{lowhigh}). For the giant magnetic M1 spin-flip resonance, we have adopted the form of a standard Lorentzian with parameterization according to~\cite{RIPL}. As can be seen from Fig.~\ref{rsfboth}, the data from the present work show an enhanced radiative strength compared to the model in the region $E_{\gamma} = 4.5 - 8.3$ MeV. In order to describe this enhancement, we have empirically added a pygmy resonance with a Gaussian form:
\begin{equation}
f_{\mathrm{pyg}} = C_{\mathrm{pyg}} \cdot\frac{1}{\sqrt{2\pi}\sigma_{\mathrm{pyg}}}\exp\left[\frac{-(E_{\gamma} - E_{\mathrm{pyg}})^{2}}{2\sigma_{\mathrm{pyg}}^2}\right].
\end{equation}
Here, $C_{\mathrm{pyg}}$ is a normalization constant, $\sigma_{\mathrm{pyg}}$ is the standard deviation, and $E_{\mathrm{pyg}}$ is the mean value (centroid) of the resonance. With the GLO E1 strength we found $C_{\mathrm{pyg}} = 4.0\cdot10^{-8}$ MeV$^{-2}$, $\sigma_{\mathrm{pyg}} = 1.5$ MeV, and $E_{\mathrm{pyg}} = 8.7$ MeV. The result is shown in the upper panel of Fig.~\ref{rsfboth}, together with ($\gamma,x$) photonuclear data from Refs.~\cite{Fultz69,Lep74,Varlamov}. It can be seen that the model description of the data works rather well, and that it is indeed necessary to include extra strength around $S_n$ in order to get a reasonable fit. It is also seen that the SLO model fits the ($\gamma,x$) data very well; however, it is not able to describe neither the shape nor the magnitude of our data below the neutron threshold.  

In the second approach we have applied results from QRPA calculations on the E1 strength~\cite{Goriely2}. It can be seen from the lower panel of Fig.~\ref{rsfboth} that the QRPA calculation tends to overestimate the E1 strength for $\gamma$-ray energies below the GEDR peak compared to experimental ($\gamma,x$) data. It is therefore probably not a correct reference for the expected E1 strength; nevertheless, it gives a lower limit on the pygmy strength. Again, we have added the M1 and the E1 strength together with a Gaussian parameterization of the extra strength around neutron threshold. As shown in Fig.~\ref{rsfboth}, the total model description of our data is not so good as in the GLO case, especially for the low-energy part. Using the QRPA E1 strength, we obtained $C_{\mathrm{pyg}} = 1.8\cdot10^{-8}$ MeV$^{-2}$, $\sigma_{\mathrm{pyg}} = 1.3$ MeV, and $E_{\mathrm{pyg}} = 8.0$ MeV for the pygmy resonance.

It is clear from our analysis that extra strength is present below and above the neutron threshold in $^{117}$Sn. Our data show the functional form of the pygmy resonance below $8$ MeV from low to high $\gamma$-ray energies for the first time. In addition, this resonance has not been experimentally measured in a stable, odd-$A$ tin isotope before, and this has now been successfully done in $^{117}$Sn.

Measuring the enhancement of our data in the energy region $E_{\gamma} = 4.5 - 8.0$ MeV relative to the GLO model plus the Lorentzian M1, we estimate an excess strength of $11 \pm 1$(stat) $\pm 3$(syst) MeV$\cdot$mb. Using the QRPA E1 strength plus the Lorentzian M1 as a reference for the expected strength function in the same energy region, an excess strength of $8 \pm 1$(stat) $\pm 3$(syst) MeV$\cdot$mb is estimated. For the total pygmy strength summed over all $\gamma$-ray energies, we find $40 \pm 15$ MeV$\cdot$mb for the GLO approach, and $17 \pm 8$ MeV$\cdot$mb using the QRPA calculation.

If one assumes that all of the excess strength is E1, then this yields 0.6(2)\% of the TRK sum rule for $ 4.5 \leq E_{\gamma} \leq 8.0$ MeV using the GLO approach. For the QRPA calculation we obtain 0.4(2)\% of the TRK sum rule for the same energy region. Correspondingly, the total pygmy strength is 2.3(8)\% and 1.0(5)\% of the TRK sum rule using the GLO E1 strength and the QRPA E1 strength, respectively. 

%As our experimental data give the total RSF, one should keep in mind that the total pygmy strength is very sensitive to the choice of a "background" represented by the models applied. Especially, the various models of the strong E1 component can give rather different results for the pygmy parameters as shown in Fig.~\ref{rsfboth}. However, we stress that the total, experimentalÊRSF is well determined by our data. It is a challenge for theory to describe and explain the experimental results with good precision, as none of the available models found in literature seem to have the necessary predictive power for a wide energy region. 

The nature of the enhancement is at present undetermined. Since the present experimental technique does not distinguish between electric and magnetic transitions, the enhanced strength could in principle be due to both E1- and M1-type radiation. However, since a large number of E1 transitions have been found in previous NRF experiments~\cite{Govaert} and a pygmy resonance of E1 character has been identified in the exotic $^{129-133}$Sn nuclei~\cite{Adrich}, it is probable that the enhancement seen in the Oslo experiment is also due to E1 radiation. It is highly desirable to firmly establish the electromagnetic character, the multipolarity, and the absolute strength of the enhancement in $^{117}$Sn utilizing other experimental techniques such as, e.g., (n,$2\gamma$) experiments.

In conclusion, the total RSF in the quasi-continuum for $^{117}$Sn up to $E_{\gamma} = 8$ MeV has been measured with the Oslo method. A significant enhancement in the strength is observed in the energy region $E_{\gamma}= 4.5-8.0$ MeV. This enhancement is compatible in strength and position with the pygmy resonance observed previously in even-even Sn nuclei. For the first time, the pygmy resonance has been measured in an odd, stable Sn nucleus, and the functional form of this resonance has been determined from low to high $\gamma$-ray energies.

\acknowledgments 

This research was sponsored by the National Nuclear Security Administration
under the Stewardship Science Academic Alliances program through DOE
Research Grant No. DE-FG52-06NA26194. 
U.~A. and G.~E.~M. also acknowledge support from U.S. Department of Energy Grant No. DE-FG02-97-ER41042. 
Financial support from the Norwegian Research Council (NFR) 
is gratefully acknowledged.

\end{document}